\documentclass[seceq,epsf]{ptptex}
\usepackage{graphicx}
\def\vec0{\mbox{\boldmath $w^0$}}

\def\ub{unit$_{\rm b}$}
\def\uf{unit$_{\rm f}$}

\title{
Realization of features of immune response by dynamical system models
and a possible mechanism of memory of antigen invasion
}

\author{Mika Yoshida, Kinji Fuchikami$^*$ and Tatsuya Uezu    
}

\inst{
Graduate School of Humanities and Sciences,\\
Nara Women's University, Nara 630-8506, Japan\\
$^*$ In-Silico Sciences, Inc. Tokyo 145-0065, Japan
}

\abst{
Among features of real immune responses which occur when
antigens invade a body,
there are two remarkable features.
One is that the amount of antibodies produced in the secondary invasion
by the same antigens is  more than 10 times larger
than that in the primary invasion.
The other is that more effective antibodies which can
neutralize the antigens more quickly are produced by somatic hypermutation
during the immune response.  This phenomenon is named as
'affinity maturation'.

In this paper, we try to reproduce these features by 
dynamical system models and present possible 
factors to realize them.  Further, we present a  model in which 
the memory of the invasion by antigens is realized without
immune memory cells.

}

\begin{document}

\maketitle

\section{Introduction}

The immune system has evolved specific mechanisms to defend against numerous
invading pathogens and any toxic molecules they produce. The acquired immune
system effectively eliminates foreign molecules (i.e., antigens) from its
body. This specific immune system can distinguish the body's own tissues
(self)  from tissues and particles not normally found in own
body (non-self), and removes the non-self antigens efficiently to protect
itself from harmful environments. 
In this paper, we propose dynamical system models which can reproduce
most characteristic phenomena in immune responses.

First, we briefly summarize specific immune responses
\cite{Illustrated98,TheCell}.
In specific immune responses, many cells and physiologically active 
molecules interact in complicated ways.
Among them, main constituents are B-lymphocytes (B-cells) produced in
the Bone marrow, T-lymphocytes (T-cells) produced in
the Thymus and antibodies (immunoglobulins).
B-cells generate and secrete antibodies.  An antibody attaches to
a specific antigen and makes it easier for phagocytes to
destroy the antigen.  T-cells attack antigens directly, 
and provide control of the immune response.
An antibody is a protein which possesses an immunoglobulin structure
consisting of a variable region (V region) and a constant region (C region).
The V region allows a variety of specific association with the corresponding
antigen. On the other hand, the variety of C region is mainly related to 
the way of the antigen elimination. According to the difference in C regions,
antibodies are classified as IgM, IgG, IgE etc. 
On the surface of each B-cell, there exists a B-cell receptor (BCR),
which is an antibody with a transmembrane structure.
Now, let us explain the V region in more detail.
The V region of a BCR (and also an antibody)
has proper  3-dimensional structure.
This consists of several  small structures which are
called idiotopes.  
A set of idiotopes can bind to the antigenic determinant of a specific antigen.
This set of idiotopes is called a paratope
and the antigenic determinant is called an epitope.
Further, an idiotope  can be recognized by other antibodies (anti-idiotypic 
antibodies).
The total set of idiotopes in the BCR is called 'Idiotype'.

A family  of B-cells which are generated from a B-cell 
are called 'clones'.
Therefore, a clone and antibodies produced by clone have the same
idiotype.
For each of various antigens,  clones with  specific idiotypes
selectively correspond and can respond to the antigen (clonal 
selection theory).
On the other hand, killer (cytotoxic) T-cells 
have a function to kill cells infected by
viruses and cancer cells  etc. (cell-mediated immunity).
Further, another type of T-cells, so called helper T-cells,
play several roles to regulate immune
responses mediated by B-cells through antigen-antibody interactions
 (humoral immunity).  Helper T-cells promote 
the maturation and the proliferation of  the B-cells responding to the
specific antigens, and terminate  immune responses by suppressing the
maturation and the proliferation of B-cells after neutralization of
antigens.

In this paper, we propose models 
in which following features 1 and 4 in real immune systems are realized,
taking into account features 2, 3, 5 and 6.

Features in real immune systems:

\begin{enumerate} 
\item
The amount of antibodies produced by the secondary immune response is
more than 10 times larger than that 
produced by the primary immune response\cite{Illustrated98}.
\item
When an antigen invades the system, a clone with high 
affinity to the antigen is selected (clonal selection theory)\cite{Illustrated98}.

\item After exposure to the antigen, class switching and somatic 
hypermutation occur in B-cells, and several kinds of antibodies
with high affinity to the antigen are produced
\cite{AID02,Song-etal98}.
\item Among B-cells which are derived by somatic hypermutation,
the B-cells with higher affinity to the antigen
remain and secrete many antibodies (affinity maturation)
\cite{TheCell,Song-etal98}.
\item 
The apoptosis (cell death) rate of  B-cells decreases 
during a  proliferation period in 
response to antigens\cite{Eeva2004}.
\item
Immune memory cells are created
during the primary immune response
 and these are reserved for as long a period as the
individual life\cite{Illustrated98}.
\item 
One immune response generates the other types of
antibodies responding to the original antibody
(existence of anti-idiotypic antibodies)\cite{Illustrated98}.
\end{enumerate}

Now, we briefly explain these features and how to include them into 
models in this paper.
\begin{enumerate}
\item[1.] 
In real systems, the amount of antibodies produced in 
the secondary response ranges from 10 to 1000 times
 as many as that in the primary response.
In this paper, as a criterion of the realization of feature 1,
 the amount of the  antigens in the secondary
response is set to 10 times the amount of antigens in the primary response.
This number 10 is rather arbitrary and it can be replaced 
by a larger multiple of 10 such as 100 etc.
\item[2.]
In a body, there are a huge number of immune cells and
they can respond to any kind of antigens.
When any antigen invades the system, there exists at least one clone 
whose affinity to the antigen is high and  clone is stimulated and 
responds to the antigen. This is the clonal selection
theory and has been confirmed experimentally.

\item[3. and 4.]
Here, we explain class switching, somatic hypermutation and affinity
maturation.
Before invasion by an antigen,
  B-cells produce antibodies of the class IgM.
These antibodies have lower affinity to neutralize the
antigen.  After the invasion of the antigen, 
depending on  the amount of the antigen and its kind,
B-cells switch the class of antibodies they produce.
In recent studies, 
 the enzyme has been found to induce class switching
and the  mutation of the V regions of the
B-cells simultaneously. The latter is called somatic hypermutation. 
Among B-cells which are created through 
 class switching and somatic hypermutation, 
the B-cells  which have higher affinity than those 
 before  class switching play a main role in the secondary
response.
In fact, it is known that the antibodies produced in the secondary
response have higher affinity to the antigen than those produced
in the primary response\cite{Illustrated98}.
This phenomenon is called affinity maturation.
Affinity maturation is considered to take place in
two processes\cite{Illustrated98}.
\begin{enumerate}
\item At the later stage of the primary immune response,
clones with higher affinities are produced.
\item In the secondary immune response, clones with higher affinities
proliferate selectively by the stimulus of the antigen.
\end{enumerate}
In this paper, we assume that 
 B-cells which are selected by the clonal selection 
in the primary response switch their class  and undergo
somatic hypermutation,
and  there appear several B-cells with lower 
and higher
affinities than  B-cells  before somatic hypermutation.
Thus, the first  of the above two processes is assumed although
it is not guaranteed that the B-cells with higher affinities dominate
in the primary response.
Therefore, our main purpose is to realize the second  of the processes.
We investigate whether the concentrations of the B-cells
 which have higher affinity become
dominant in the primary and the secondary responses.
\item[5.]
Usually, on their genes, cells are programed to die naturally. 
This spontaneous death of the cells is called
apoptosis.
However, it has been reported that for maturated B-cells
apoptosis is suppressed during a proliferation period.
The experiment to prove this phenomenon is as follows.
First,  an appropriate  amount of stimulus is given.
Then, the amount of B-cells increases.  After that,
 if the amount of the stimulus is either increased or decreased,
the amount of the B-cells always decreases.
That is, in the beginning of the immune response when B-cells 
proliferate, apoptosis of the B-cells is suppressed. 
In our model, the stimulus is expressed by 
the sensitivity $\sigma$ which is the product
of the affinity to an antigen and the concentration of the antigen.
When the antigen invades the system,  $\sigma$ is decreasing
monotonically, because we assume that after an amount of 
the antigens invade the system  they do not proliferate but just
are neutralized.  Therefore, to include feature 5, we assume that if 
$\sigma$ is large and the maturation rate and the proliferation
rate are high, the apoptosis rate of the B-cells is low, and it is  high
	  otherwise.
\item[6.]
B-cells created by somatic hypermutation finally disappear
after the immune response because of apoptosis.
However, a fraction of B-cells become immune memory cells by
differentiation and their lifetimes are as long as those of
the living body.  In the secondary response, 
these memory cells are activated, become
activated memory cells, and  secrete antibodies with IgG class.
Therefore, antigens are neutralized efficiently from the
beginning of the secondary response.
To include these facts into model, we assume
the following in the present paper.\\
We reserve the part of the concentration of each clone
as the immune memory cells 
every time its concentration increases by some constant amount.
We assume that the apoptosis rate of the immune memory cells is
0 when they are produced, and are not activated in the primary response.
In the secondary response, we assume that the
immune memory cells are activated when the  amount of antigens 
exceeds some threshold value, and activated memory cells 
 have a finite apoptosis rate but
have different response properties to the antigens from the normal
  B-cells.
\item[7.]
It is known that in a real system, there exist anti-idiotypic
antibodies which can interact with an antibody\cite{Illustrated98}.
In the last part of this paper, we study a model in which both the antibodies
and the anti-idiotypic antibodies are taken into account.
By this model, the memory of the invasion by the antigens
without immune memory cells is studied.
\end{enumerate} 

To realize features 1 and 4, we study second generation immune
network model which was introduced by Varela et al\cite{Varela91}.
Although this model is based on the idiotypic immune network theory
proposed by Jerne\cite{Jerne74},  we do not consider the interaction 
between immune cells except for last model (model 3),
 but consider only the interaction between antigens
and immune cells. That is, we do not take a network point of view.

First, we study a model taking into account features 2, 3, 5 and 6.  
Model is described by an ordinary differential equation system. 
We call it model 1.
By this model, we can realize feature 4 when the amount of antigens is
in a certain range.  However, the sum of the concentrations of antibodies
produced in the secondary response is only  several times as 
many as that produced in the primary response.
That is, feature 1 is not realized.

In real systems, it takes time for 
the B-cells to recognize the antigens
when they invade a system.  That is, there exists
the time delay for the B-cells to recognize the antigens.
Taking the time delay into account, we study a modified model,
which is now described by a delay-differential equation system.
We call it model 2.
Then, we can realize both features 1 and 4. Therefore, it is 
clarified that the time delay is one of 
the most important factors to realize
feature 1.

The above results are obtained by taking into account feature 6,
the existence of immune memory cells.
It is very interesting to study whether the concentrations of 
antibodies which can respond to antigens are retained
spontaneously  without the immune memory cells, 
even after the primary response finishes.
In other words, we have interest in another mechanism of memorizing the
invasion by the antigens without assuming the immune memory cells.
In order to search for this possibility,
 we study  a model assuming feature 7, the existence of
antibodies and  corresponding anti-idiotypic antibodies,
without the immune memory cells.  We call it model 3.
As a result, we find that several pairs of  antibodies
and anti-idiotypic antibodies are excited and
 their concentrations continue to oscillate 
even after the primary response finishes, although
 affinity maturation does not necessarily
take place.

The construction of the paper is as follows.

In section 2, we explain 
basic model. Models 1, 2 and 3 are studied in sections 3, 4 and
5, respectively. We give summary and discussions in section 6.

\section{Basic model}

As basic model, we use second generation immune network model
which was introduced by Varela et al.\cite{Varela91}.
The main constituents of Varela model are 
B-cells, T-cells and free antibodies produced by B-cells.
The role of the T-cells is taken into account through the
interaction between B-cells and antibodies in present model.
 Let us distinguish idiotypes by index $i$.
 Let us call the  B-cells with $i$-th idiotype clone $i$,
and denote   their concentration by $b_{i}$
 and the concentration  of the  free antibodies  produced
by the B-cells by $f_{i}$.
The sensitivity of the network for the $i$-th idiotype
is defined as follows;
\begin{eqnarray}
\sigma_{i}=\mathop{\sum}_{j=1}^{N}m_{ij}f_{j},\;(m_{ii}=0)
\label{eq:sigma}
\end{eqnarray}\\
where $N$ is the number of idiotypes.
 $m_{ij}$ is called the affinity, which represents the
strength for the B-cells (and T-cells)  with the $i$-th idiotype
to detect the  antibodies with the $j$-th idiotype.
This is a general setting for  immune network model.
In this paper, we do not take a network point of view.
Thus, instead of (\ref{eq:sigma}), we define $\sigma _i$
as 
\begin{eqnarray}
\sigma_{i}=m_{iA}A.
\label{eq:sigma2}
\end{eqnarray}
Here, $A$ is the concentration of the antigen which invades the system
 and $m_{iA}$ is the strength for clone $i$ to detect
the antigen.\\
The number of B-cells and antibodies
change in time by the following causes.
Free antibodies
are removed from the system 
 because they have a natural lifetime
and also they interact with antigens and 
neutralize them.  On the other hand, 
they are produced by B-cells as a result of the
 maturation of B-cells.
The probability of the maturation is assumed to depend
on their sensitivity $\sigma$.
This effect is expressed by the maturation function $M(\sigma)$.

In a real immune system, if the amount of molecules is very many,
 the system regards the molecules as a part of the self and
does not respond to them.  To reflect this fact, we assume 
that when $\sigma$  representing the amount of stimulus is large
enough, the maturation does not take place and we set $M(\sigma)=0$.
Around the boundary value of the stimulus that the system can respond
 to, we assume that $M(\sigma)$  decreases as $\sigma$ increases.
When the antigens are almost removed from the system and the
immune response comes to an end, the creation of antibodies is 
 suppressed.  It is considered that
the maturation does not take place
when the amount of the stimulus becomes small.
  Thus, $M(\sigma)$ should decrease toward 0 as
$\sigma$ decreases toward  0 and
 is assumed to have a convex profile illustrated in  Fig.1.
\begin{figure}[ht]
\centering
\includegraphics[width=0.5\textwidth]{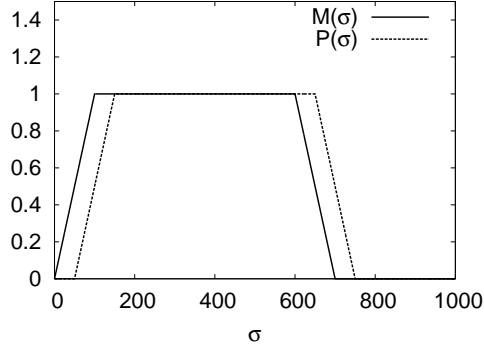}
\caption{Maturation and proliferation functions
for normal cells , $M(\sigma)$ and $P(\sigma)$.}
\label{fig:1}
\end{figure}
Correspondingly B-cells carrying $i$-th idiotype
 on their surfaces decay at a given rate 
because of apoptosis and they proliferate when they maturate.
The probability of the proliferation of B-cells 
is represented by the proliferation function $P(\sigma)$.
The B-cells do not proliferate by the stimulus of the self 
as in the case of the maturation.
When the stimulus becomes weak and the immune response comes to an end, 
the proliferation of B-cells is suppressed by T-cells. 
Therefore, we assume that $P(\sigma)$ decreases 
as $\sigma$ decreases for small $\sigma$.
Thus, $P(\sigma)$ also has a convex shape.
Further, it is considered that the suppression of the proliferation
begins before that of the maturation does.
Thus, it is reasonable to assume
that $P(\sigma)$ is shifted to right from $M(\sigma)$ (Fig.1).
In present model, we assume that $M(\sigma)$ and $P(\sigma)$
have the same shape of a symmetric trapezoid. Its slope
is $a$ and its length of the smaller size is $d$.
The length of the shift between  $P(\sigma)$ and $M(\sigma)$ is $h$.
We set $a=0.01, d=500$ and $h=50$.

Then, the time evolutions  of $f_i$ and $b_i$ are given by
the following ordinary differential equation system.
\begin{eqnarray}
\frac{df_{i}}{dt} &=&
-K_{1}\sigma_{i}f_{i}-K_{2}f_{i}+K_{3}M
\left(\sigma_{i}\right)b_{i},
\label{eqn:F}\\
\frac{db_{i}}{dt} &=&
-K_{4}b_{i}+K_{5}P\left(\sigma_{i}\right)b_{i}+K_{6},
\label{eqn:B}
\end{eqnarray}
where $K_{1}$ is the rate of the antigen neutralization,
$K_{2}$ is the rate of the death of the antibody,
 $K_{3}$ is the rate of
 the creation of the antibodies by B-cells, 
 $K_{4}$ is the apoptosis rate of the B-cells
 and $K_{5}$ is the rate of production
 of the B-cells.  Further, the term $K_{6}$
is added to take into account
the cells that are recruited into the active network
from the bone marrow.
As these parameters, we adopt the following values which
are estimated from real data.
$K_1=0.001$[day$^{-1}$ unit$_{\rm f}^{-1}$ ],
$K_2=0.15$[day$^{-1}$],
$K_3=1.0$[ unit$_{\rm f}$ unit$_{\rm b}^{-1}$ day$^{-1}$ ],
$K_{4}=0.5$[day$^{-1}$],
$K_5=1.5$ [day$^{-1}$], $K_6=0.1$ [ unit$_{\rm b}$ day$^{-1}$].
Here, 
1 unit$_{\rm b}$ is the amount of B-cells with an idiotype
which is supplied by the Bone marrow in 10 days, 
and 1 unit$_{\rm f}$ is the amount of antibodies which is
produced by the B-cells with an idiotype per day.

This is basic model. Based on this model, we consider
several models taking into account features listed in $\S1$.

In the next section,  we study model 1 in which features 2, 3, 5
and 6
are taken into account.

\section{Introduction of somatic hypermutation, change of apoptosis rate and immune
memory cells: Model 1}
Here, we explain first model, model 1.  See Fig.2.
\begin{figure}[ht]
\centering
\includegraphics[width=0.5\textwidth]{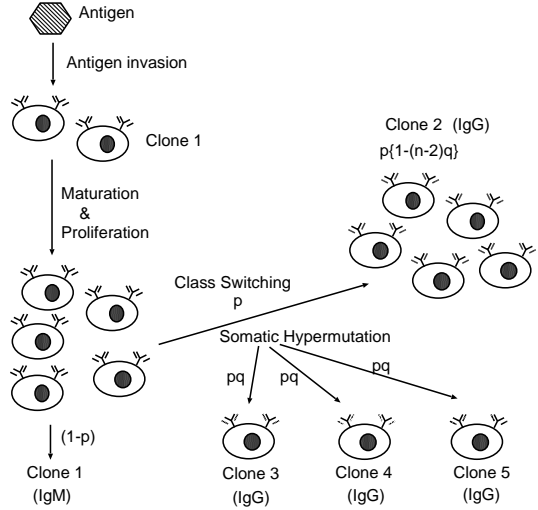}
\caption{Explanation of model 1 in which somatic hypermutation takes
 place. Symbols in the parentheses denote the classes of secreting  type of immunoglobulins.}
\label{fig:2}
\end{figure} 
We assume the following.
Let us consider the situation that one kind of antigen invades the system.
Let $A$ be the concentration of the antigens.
By the clonal selection, among several B-cells which can detect the antigens,
some clone  which detects the antigens most effectively is selected.
Let this clone be clone 1 and let 
$ m_{1A}$ be the strength of the affinity for clone 1 to detect the antigens. 
We set $m_{A1}=m_{1A}=m_1$ for simplicity.
The parameters $K_1$ to $K_6$ for clone 1 are equal to those
in basic model.

Initially, clone 1 is in the rest state,
that is, $\displaystyle b_1=\frac{K_6}{K_4}$ and $f_1=0$,
the stationary state of eqs. (\ref{eqn:F}) and (\ref{eqn:B}).
During the first invasion by the antigens,  when $b_1$ exceeds
30[unit$_{\rm b}$], class switching and somatic hypermutation
take place at the same time.  
For simplicity, we assume that somatic hypermutation takes place only once.
Further, we assume that clone 2
is produced by class switching, and clone 3 to
clone $n$ are produced by  somatic hypermutation.  Here, we consider the case $n=5$.
The concentrations of clones 1 to 5 after class switching
are assumed as follows.
\[
 b_1=(1-p)b_1 ^0, \;   b_2=p\{ 1-(n-2)q \} b_1 ^0, \;
  b_3=b_4=b_5= pq b_1 ^0, 
\]
where $b_1 ^0$ is the concentration of clone 1 just before
class switching.  We set $p=0.7, q=0.1/3$.  
That is, class-switched clone 2 is 70 $\%$ of clone 1 
and 10 $\%$ of class-switched clone 2 undergoes somatic hypermutation.
We assume that clone $i (i=2,\cdots 5)$ has 
the strength of the interaction with the antigens
$m_{iA}=m_{Ai}=m_i$.
We put $m_1=2, m_2=2, m_3=3, m_4=8, m_5=1$.
The parameters $K$s for clones 2 to 5 are the same as those for
clone 1, except for $K_6$, which is set to 0.
In basic model, $K_4$ is fixed to 0.5[day$^{-1}$].
To include feature 5, we assume that 
 for clones 1 to 5 $K_4$ changes
depending on the sensitivity $\sigma$ as,
\begin{eqnarray}
K_4 & = & K_{ 4 l}  \mbox{ for }  \sigma_i \ge 50[{\rm unit}_{\rm f}],
\nonumber\\
 & = & K_{ 4 s}  \mbox{ for }  \sigma_i < 50[{\rm unit}_{\rm f}],
\label{eq:lifetime}
\end{eqnarray}
 where $K_{4l}=0.001$[day$^{-1}$] and $K_{4s}=0.5$[day$^{-1}$].

The equations of $f_i$ and $b_i$ are given by eqs. (\ref{eqn:F})
and (\ref{eqn:B}). In these equations $\sigma_i$ is given as follows.
\begin{eqnarray}
\sigma_i &=& m_{iA}A = m_i A.
\label{eqn:sigma}
\end{eqnarray}
Before somatic hypermutation, only $i=1$ is considered.
On the other hand, the equation of $A$ is given by
\begin{eqnarray}
\frac{d A}{dt} &=&
-K_{1}\sigma_{A}A,
\label{eqn:A}\\
&& \sigma_A= m_{A1} f_1 = m_1 f_1 \mbox{ (before somatic hypermutation)},\nonumber\\
&& \sigma_A=\sum_{j=1}^5 m_{Aj}f_j =\sum_{j=1}^5 m_j f_j 
 \mbox{ (after somatic hypermutation)}.\nonumber
\end{eqnarray}
Further, we assume that for clones 2 to 5,
the 0.1[\ub] amount of B-cells is transformed into  the immune memory cells 
every time its concentration increases by 25[\ub].
The immune memory cells have an infinite lifetime but are
not activated in the primary response. 

When the same type of antigens invade the system secondly,
we assume the following.
When the amount of the antigens exceed 50[\uf],
the immune memory cells with the idiotype $i$ are activated,
and they are going to have the same properties as clone $i$
except the maturation and the proliferation functions.
Here, $i =2,\cdots,5$.  
In particular, these cells reserve immune memory cells 
and the value of $K_4$ changes as (\ref{eq:lifetime}).

Since the response by the activated memory cells will be quicker
than normal cells, we assume that the 
maturation and the proliferation functions
$M(\sigma)$ and $P(\sigma)$ for activated memory cells 
vary more rapidly as functions of $\sigma$ than those for normal cells.
See Fig.3.
\begin{figure}[ht]
\centering
\includegraphics[width=0.5\textwidth]{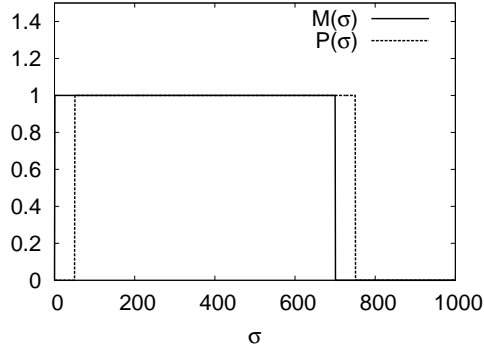}
\caption{Maturation and proliferation functions
for activated memory  cells , $M(\sigma)$ and $P(\sigma)$.}
\label{fig:3}
\end{figure}
In this paper, we assume that $M(\sigma)$ and $P(\sigma)$ have
the same rectangular shape.  Its height is 1 and
the width is 700.  $P(\sigma)$ is shifted to right from $M(\sigma)$
by $50$.  
Under these assumptions, we perform numerical simulations.

\subsection{Numerical results of model 1}

Before invasion by the antigens, 
none of the clones is activated, that is, $f_i=0$  and 
$b_i=\displaystyle \frac{K_6}{K_4}$.
We assume that at time $t=0$,
 antigens with the concentration $A_0$
invade the system and when $b_1$ reaches 30[\ub]
for the first time, somatic hypermutation takes place.
When the immune response ends, clones 2 to 5 disappear because of
$K_6=0$. 
The second invasion by the same antigens with the concentration $A_0 '$
takes place at about 100[day].

We display an example of the simulations with $A_0 =50$
and $A_0 ' =50$ in Fig.4.
\begin{figure}[ht]
\centering
\includegraphics[width=0.5\textwidth]{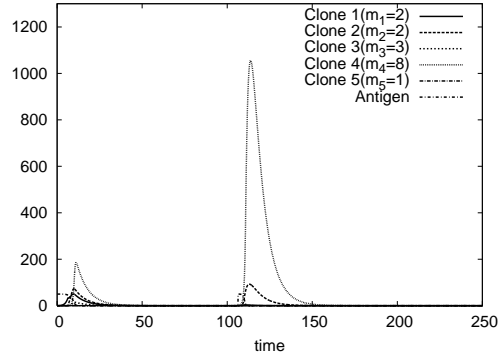}
\caption{Time series of the concentrations of antibodies $f_i$ 
 for model 1. $A_0 = 50, A_0 ' = 50 $.}
\label{fig:4}
\end{figure}
The half-life of the antigens, i.e., the period that the amount of  the initial antigens
becomes half, is 9.2 days in the primary response and is
3.8 days in the secondary response.  Thus,
the system responds to the antigens more quickly
in the secondary response than  in the primary response.
From the figure, we note that in the first and
the second invasion by the antigens
clone 4 which has the strongest interaction with the antigens
among clones  responds to the antigens mostly.
That is, feature 2, affinity maturation, is realized.
However, the sum of concentrations of all antibodies
in the secondary response is at most three times
as many as that in the primary response.
That is, feature 1 is not realized in present model.

 Next, we display another example of the simulations with 
$A_0=120$ and $A_0 ' =50$ in Fig.5.
\begin{figure}[ht]
\centering
\includegraphics[width=0.5\textwidth]{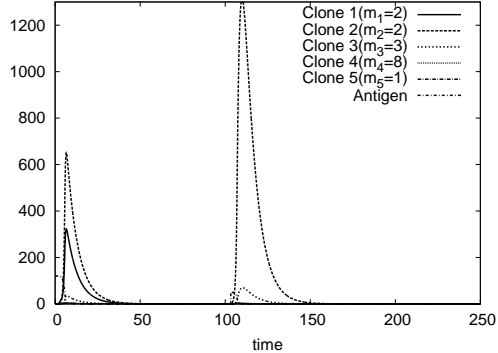}
\caption{Time series of the concentrations of antibodies $f_i$ 
 for model 1. $A_0 = 120, A_0 ' = 50 $.}
\label{fig:5}
\end{figure}
In this case, clone 2 which appears by 
class switching  responds
to the antigens mostly in the first 
and the second invasions and 
 affinity maturation does not take place.
The reason for this is considered  as follows.
For any clone $i$ with a large affinity $m_i$ to the antigens,
 if $A_0$ is large,  $\sigma_i(=m_iA_0)$ becomes very large and then
the values of $M(\sigma_i)$ and $P(\sigma_i)$ become nearly 0.
Thus, clone $i$ responds to the antigens very slowly or does not
respond to them any more.
In real immune responses, if the amount of the antigens is low,
the B-cells with high affinity respond to the antigens,
and if it is high, any B-cells respond to the antigens irrespective of their
affinity \cite{Illustrated98}.
Therefore, the present result that affinity maturation  
takes place for the concentration of the antigens which are not too small
and not too many is consistent with this fact.
The half-life of the antigens is 5.5 days and 3.7 days for the primary
and the secondary responses, respectively.
Comparing this result with that for $A_0=50$ and $A_0'=50$, we note that
the half-life in the primary response is reduced by about 4 days.
 This result also indicates that when the amount of the antigens is large,
the neutralization of antigens by
antibodies secreted by any B-cells is preferential to affinity
maturation.

In order to confirm the necessity of feature 5,
we performed a simulation in which the apoptosis rate $K_4$ is not
changed
throughout immune responses.
The result for $A_0=50$ and $A_0'=50$ is shown in Fig.6.
As is clearly seen from the figure, enough antibodies to
neutralize antigens completely are not produced, and antigens remain
after the primary response finishes.
Therefore, the reduction of the apoptosis rate in immune responses is
necessary in order to produce enough antibodies to neutralize
all antigens.
\begin{figure}[ht]
\centering
\includegraphics[width=0.5\textwidth]{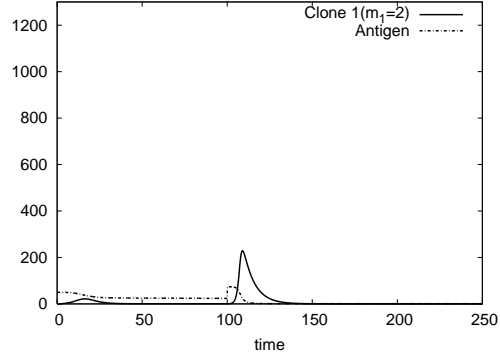}
\caption{Time series of the concentrations of antibodies $f_i$
 for model 1 in which the apoptosis rate $K_4$ is not reduced but constant.
 $A_0 = 50, A_0 ' = 50 $.}
\label{fig:6}
\end{figure}

In model 1, feature 1 was not realized.  Then,
we study model 2 taking the time delay of the response into 
account in the next section.

\section{Introduction of time delay: Model 2}

We introduce the time delay  in order to reflect 
feature that in real immune responses
the maturation and the  proliferation
for B-cells  require some time delay after exposure to
the antigens.
  This effect is realized by
substituting $\sigma _i ^m \equiv m_i A(t - \tau _m)$ 
for $\sigma _i$ in $M(\sigma _i)$
of eq. (\ref{eqn:F}) and by substituting 
$\sigma _i ^p \equiv m_i A(t - \tau _p)$ for $\sigma _i$ in $P(\sigma _i)$
of eq. (\ref{eqn:B}).  
We set $\tau_m=1.3$[day] and $\tau_p=2$[day].
Other $\sigma _i$s and $\sigma _A$ in
eqs. (\ref{eqn:F}), (\ref{eqn:B}) and (\ref{eqn:A}) are not changed.
The existence of the immune memory cells is taken into account in 
model 2 as well as model 1.
We display an example of the simulations with $A_0 =50$ 
and $A_0 ' =50$ in Fig.7.  In Fig.8, we display the semi-log plot of 
the concentrations of the antigens, IgM (clone 1) and IgGs (clones 2 to 5).
\begin{figure}[ht]
\centering
\includegraphics[width=0.5\textwidth]{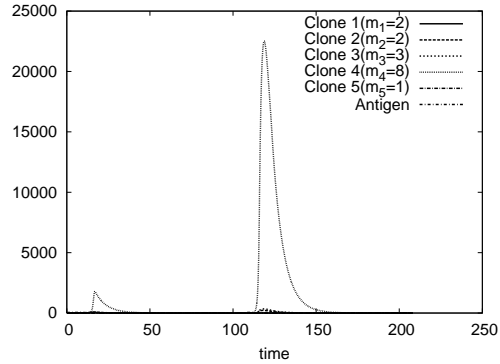}
\caption{Time series of the concentrations of antibodies $f_i$ 
 for model 2. $A_0 = 50, A_0 ' = 50 $.}
\label{fig:7}
\end{figure}
\begin{figure}[ht]
\centering
\includegraphics[width=0.5\textwidth]{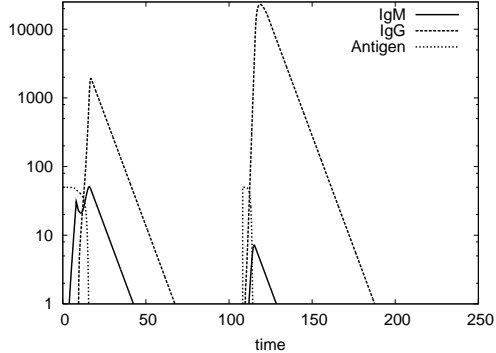}
\caption{Time series of the concentrations of antibodies $f_i$ 
 for model 2. $A_0 = 50, A_0 ' = 50 $. Semi-log plot.}
\label{fig:8}
\end{figure}
As is clearly seen from the Fig.8, 
the sum of concentrations of all antibodies
in the secondary response is more than 10 times
higher than  that in the primary response.
That is, the introduction of the time delay results in  
the creation of the huge amount of the antibodies in the secondary
response and feature 1 is realized.
Affinity maturation  is also realized.\\
Let us look at  the detail of the time sequences.
In the primary response, somatic hypermutation takes place at 9.5-th day
after the first invasion by the antigens, and the sum of IgG concentrations takes 
its maximum value 1,900[\uf]  at 17-th day.   
On the other hand, in the secondary response,
the IgM concentration becomes
 maximum  at 7-th day after the second invasion by the antigens and 
the sum of IgG concentrations takes its 
 maximum value 23,000[\uf]  at 10.5-th day.   
In real systems, the amount of antibodies becomes maximum in about
10 days to two weeks in the primary response and it becomes maximum
faster in the secondary response.  
Thus, the present results about dates when the
amount of antibodies becomes maximum are similar to those in real systems.
The half-life of the antigens is 13.7 days in the primary response and
is 5 days in the secondary response. Both of them are longer than those
in model 1, respectively.  
This is because it takes time in present model
for antibodies to increase due to the existence of the time delay 
for the response.

Further, we studied model changing the values of  
the time delays $\tau^m$ and $\tau^p$ and found 
that for $\tau ^p \ge \tau^m$, 
the amount of the antibodies in the secondary response can be
extremely high compared to those in the primary response.
On the other hand, for $\tau ^p <\tau^m$, 
the amounts of antibodies are similar both in the primary and secondary
responses.  We investigated antibody concentrations by changing
$\tau^p$ under the condition $\tau ^m \ge \tau^p$.
Then, we found that at $\tau^p \ge 2.6$, the concentration of an
antibody goes to infinity  in the secondary response.
This implies that in present model the antibodies in the secondary response
can be any amount if $\tau^p$ and $\tau^m$ are  chosen appropriately,
and the criterion on the amount of antigens
 for feature 1 we set is irrelevant to
the conclusion.
 Therefore, we could construct model in which 
features 1
and 4 are satisfied by taking into account features 2, 3, 5 and 6.

\section{Introduction of anti-idiotypic antibody : Model 3}

In the previous sections, we have assumed the immune memory cells.
In  real systems, it is known that when one kind of antigens invade the system,
not only  the antibodies which interact with the antigens 
are produced, but also anti-idiotypic antibodies which 
interact with the antibodies are produced.
Therefore, theoretically, it is possible that 
the invasion by the antigens provokes 
the creation of the antibodies and the anti-idiotypic antibodies, 
and they stimulate each other
and retain their concentrations spontaneously, 
even after the elimination of the antigens.

In order to investigate whether this scenario is possible
or not, we introduce an anti-idiotypic clone C$_1$ which can respond to
clone 1, 
and also a clone C$_i$ which can respond to 
 clone $i$ that appears by somatic hypermutation.
For simplicity, we do not consider  class switching.
Since we mainly have interest in the possibility of retaining 
the concentrations of immune cells which are produced after
the first invasion by an antigen,  we do not assume the immune memory
cells and the change of the apoptosis rate, and do not introduce the time delay.
That is, model 3 is basic model with somatic hypermutation
and anti-idiotypic antibodies.

Now, we explain model 3 in detail. See Fig.9.

\begin{figure}[ht]
\centering
\includegraphics[width=0.5\textwidth]{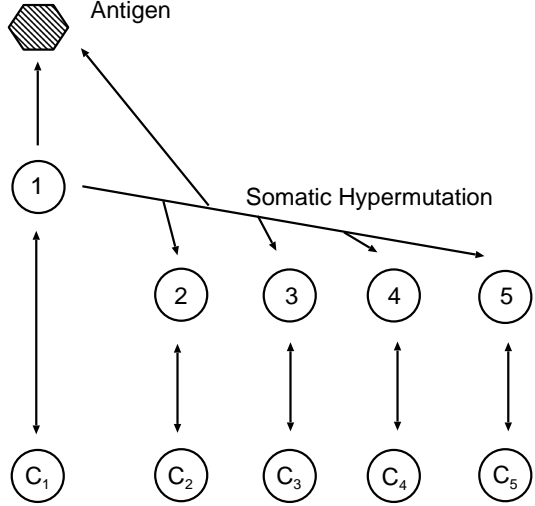}
\caption{Explanation of model 3 in which the antibodies and
anti-idiotypic antibodies are taken into account.}
\label{fig:9}
\end{figure}

We assume that clones 2, $\cdots$, 5 appear by somatic hypermutation.
Clones 1, $C_1, \cdots, C_5$ have the same parameters $K_1, \cdots,
K_6$ as in basic model.  On the other hand,
for clones 2 to 5, the  
 parameters $K_1, \cdots,K_5$ are the same as in basic model,
but $K_6=0$.
  Initially, clone 1, and clones $C_1$ to $C_5$ are in the rest state.
We assume that the strength of the interaction $m_{i C_i}=m_{C_i i}$
between clone $i$ and clone $C_i$ is 10 for $i=1, \cdots, 5$.
As for the strength of the affinity $m_{iA}=m_{Ai}=m_i$, we set 
$m_1=2, m_2=2.5, m_3=1, m_4=8$ and  $m_5=10$.
 Further, we adopt a more natural
assumption that somatic hypermutations take place at different times
successively, because all phases of  the oscillations would be the same
if they took place at the same time as in models 1 and 2.
We assume that at the first invasion 
by the antigens, when $b_i$ exceeds 30[\ub] for the first time, 
 clone 1 undergoes somatic hypermutation and clone 2 appears.
Further, we assume that somatic hypermutation takes place three times,
once a day after the first somatic hypermutation.  
At each somatic hypermutation, 20$\%$ of the concentration
of clone 1 is transformed into  new clone.
That is, we put $b_1=0.8b_1 ^0$ and $b_i=0.2b_1 ^0 (i=2 - 5 )$,
where $b_i ^0$ is the concentration of clone 1 just before the
somatic hypermutation.  Under these assumptions, we perform numerical 
simulations.

We display two examples of the simulations with $A_0 =55$ 
and $A_0 ' =55$ in Fig.10 and  Fig.11.
In two examples, the timing of the second invasion by
the antigens differs.
\begin{figure}[ht]
\centering
\includegraphics[width=0.5\textwidth]{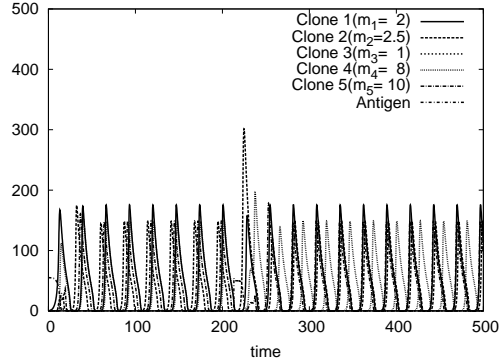}
\caption{Time series of the concentrations of antibodies $f_i$ 
 for model 3. $A_0 = 55, A_0 ' = 55$.}
\label{fig:10}
\end{figure}
\begin{figure}[ht]
\centering
\includegraphics[width=0.5\textwidth]{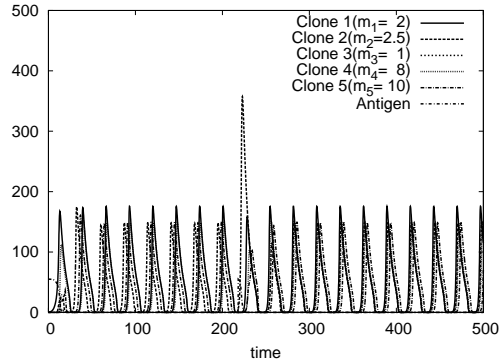}
\caption{Time series of the concentrations of antibodies $f_i$ 
 for model 3. $A_0 = 55, A_0 ' = 55$.
The timing of the second invasion is different from that in Fig.10.}
\label{fig:11}
\end{figure}
As is seen in Fig.10, after the first invasion, 4 pairs of clones
 remain showing oscillatory behavior, but clone 3
which has the smallest $m$ disappears.
The similar results that clones with the smaller $m$ disappear
 are obtained for other choices of $m$s, if
the values of $m$s are not too large.
That is, the first process of affinity maturation,
which is explained in the introduction, is realized.
On the other hand, if the value of $A_0$ is large, 
clones with smaller $m$ remain and the others disappear.
In the secondary invasion, remaining clones are not necessarily
clones with larger $m$.  In fact,
whether a clone remains or not depends on the phase of the 
oscillation of the antibody  concentration $f_i$ at the moment when the
antigen invades.  
If $f_i$ is small when the antigen invades, $f_{C_i}$ is large  and then 
$\sigma_i(=m_iA+m_{iC_{i}}f_{C_i})$ is large.
Thus, $f_i$ and $b_i$ tend to 0.  For $i \ge 2$, once 
 $b_i$ tends to 0, clone $i$ disappears, because of $K_6=0$,
i.e., no B-cell supply from the Bone marrow. 
However, since the phase of the oscillation of each clone differs from
 each other,
 not all clones disappear.
On the other hand, feature 1 was not realized in this model.

\section{Summary and discussions}

We studied the second generation immune network model introduced 
by Varela et al., and taking into account features of the immune
response we tried to realize the following two observed features.

1. The amount of antibodies produced by the secondary  response is
more than 10 times larger than that produced by the primary response\cite{Illustrated98}.

4. Affinity maturation. 
Among B-cells which are produced by somatic hypermutation, B-cells with
     higher affinity to antigens remain, and they secrete a huge amount of
     antibodies\cite{TheCell,Song-etal98}.

First, we studied model 1 in which somatic hypermutation of clones, 
 the immune memory cells and the change of 
the apoptosis rate are taken into account.
Then, we found that affinity maturation is realized
for some range of the concentration of invading antigens.
This result is consistent with the known fact 
that  if the amount of the antigens is low,
the B-cells with high affinity respond to the antigens,
and if it is high, any B-cells respond to the antigens irrespective of their
affinity.

As for immune memory cells, we assumed that they are activated when the
antigen concentration exceeds a threshold in the secondary response.
The reason for setting the threshold of the activation of the memory
cells not for the sensitivity but for the antigen concentration 
is to see its effect by activating all memory cells at the same time.
If we gave the threshold of the activation of the memory cells for
$\sigma$, affinity maturation would be realized more clearly. 

On the other hand, we could not realize feature 1
in model 1.  
The reason is considered as follows.
In model 1, when the antigens invade the system secondly,
the immune memory cells become the activated B-cells and their
maturation takes place instantaneously.
Thus, the system neutralizes the antigens before the proliferation
of the B-cells takes place sufficiently.
Then, the concentrations of antibodies do not increase very much.
In reality, when the antigens invade  the system,
it takes time for the immune cells to recognize the antigens.
Thus, it is natural to introduce the time delay for 
the system to detect the antigens.  
Therefore, as a more realistic model, we considered model 2
in which the delay time $\tau$ of the response to the invasion
by the antigens is taken into account.
We assumed somatic hypermutation of B-cells, the immune memory cells
 and the change of the apoptosis rate as well as  in model 1.
As a result, the response of the system at time $t$ is caused
by the amount of the antigens at time $t - \tau $,
and it takes time for the immune memory cells to maturate, 
and then the B-cells have enough time to 
proliferate and feature 1 is realized.
Thus, we found that the time delay is one of the most important factors 
to realize feature 1 in present model.
Another important factor is the introduction of different maturation and
proliferation functions for activated memory cells from those for normal
B-cells. 
If we adopt the same maturation and proliferation functions for
activated memory cells as those for normal B-cells, the amounts of
antibodies do not differ between the primary and secondary responses
even if the time delay is taken into account.
This assumption to take different maturation and proliferation functions
in normal and activated memory cells is reasonable, because in
reality, the activated memory cells respond to antigens more quickly
than normal cells.

In models 1 and 2, as a mechanism of memorizing the 
primary invasion by antigens, 
we assumed the existence of immune memory cells
whose maturation and proliferation functions increase rapidly
for small value of the sensitivity $\sigma$.
It is interesting to see if there is another mechanism
to memorize the invasion by antigens without assuming
immune memory cells.
One such candidate is the interaction between
antibodies and their anti-idiotypic antibodies 
 which was originally proposed by
 Jerne in order to activate an immune system spontaneously 
without the stimulation by antigens.
In order to see whether the idea works or not, in model 3, we introduced
anti-idiotypic antibodies which respond to antibodies.
We did not assume the immune memory cells, the change of the apoptosis
rate and the time delay in the immune response.
As a result, we found that the several pairs of antibodies and
anti-idiotypic antibodies with higher affinities
stimulate and inhibit each other,
and their concentrations oscillate in time and are retained
after the primary response finishes.
Thus,  the first process (a) of affinity maturation is realized.
Although several clones always survive after the secondary invasion by the
antigens, clones which can survive  depend
on the timing of the invasion because of the oscillatory nature of 
the concentration of each clone.  Clones with higher affinity 
 do not always survive.
Further, we found that feature 1 is not realized.
We surmise that this is because the immune memory cells, the time delay
and the change of the apoptosis rate are not included in this model.
Although  model 3 does not satisfy the desired features 1 and 2,
the mechanism to retain the concentration of  
the immune cells in this model is interesting 
as  the  possibility of the memory 
of the invasion by the antigens other than the immune memory cells.

Here, we  make several comments on the values of parameters 
 we adopted in the paper.\\
As for $K_1$ to $K_6$, $\tau ^m $ and  $\tau ^p $,
realistic values were chosen.  As for other parameters such
as $K_{4l},  A_0, A_0 '$, the affinities $m$s, and the maturation 
and the proliferation functions,
we tried a lot of choices.
Now, let us discuss  $M(\sigma)$ and $P(\sigma)$.
We adopted the shape of a symmetric trapezoid for both functions
of normal cells.
One reason  for this was  the simplicity and another was that
we wanted to study the effect of sensitivity $\sigma$ because the
$\sigma$ dependences of these functions are simple and 
easy to identify the regions such as
the fully maturated region, the region corresponding to 
 the self recognition, and so on.
We assumed that $P(\sigma)$ is shifted to right from $M(\sigma)$  by
$h$.
This seems reasonable, because it is considered that the suppression of the proliferation
begins before that of the maturation does.
The value $h$ is set to 50 in the paper.
If the shift $h$ is too large, the proliferation ends soon,
and it takes time to neutralize the antigen.
$M(\sigma)$ is  assumed to be proportional to $\sigma$ in
the interval $\displaystyle[0, \frac{1}{a}]$.
That is, the system responds to the antigens even if their
concentration is very small.  However, this assumption is not necessary.
We tried other $M(\sigma)$ such that   $M(\sigma)=0$  
in the interval $\displaystyle[0, \sigma_0]$ for some $\sigma_0>0$.
If $\sigma_0$ is not too large, it does not take much time to neutralize
the antigen.\\
The important parameters in $M(\sigma)$ and $P(\sigma)$ are 
the slope $a$ of the trapezoid and the length $d$ of
its smaller size.
Roughly speaking, the typical values of the antigen
concentration $A^*$ and the affinity $m^*$ to the antigen
are determined by the relation
\begin{eqnarray}
\frac{1}{a}+ \frac{d}{2}=m^* A^*.
\label{eq:typical}
\end{eqnarray}
The left hand side of eq. (\ref{eq:typical}) is nothing but
the central point of the interval of the sensitivity $\sigma$
in which $M(\sigma)$ is maximum.
This relation is interpreted in terms of affinity maturation.
That is, how much affinity maturation there is, the degree of affinity maturation,
depends on the amount of the antigen. The affinity, with which B-cells
can respond to the antigen, is
large for the small amount of the antigen,
and small for the large amount of them.\\
As for the apoptosis, if we do not assume this,
that is, if we fix $K_4$, the antibodies can not
increase sufficiently and the antigens are not completely neutralized as
was shown in this paper.
Therefore, the reduction of the apoptosis rate
 is also one of the important factors
to realize features 1 and 2.

The  values of parameters used in the paper were
chosen by trial and error so that observed features are realized.
We also confirmed that the robustness of the values of these parameters, 
that is, slight change of the values of
the parameters did not change the qualitative behavior
of the system. For example, if we take similar maturation and
proliferation functions in both  normal and activated memory cells, and make the
height of those functions larger in activated memory cells than in normal
cells, we get a lot
of antibodies in the secondary response.
Further, we investigated immune responses by changing other parameters
of maturation and proliferation functions for activated memory cells.
As a result, we found that it is possible for activated memory cells to
produce a lot of antibodies by making the slope of the left edge of the 
trapezoid steeper.

As for the plausibility of the
values of the parameters, we have not looked into it in this paper,
because we do not have interest in constructing realistic models 
taking into account as many factors in real systems as possible,
but have interest in the mechanisms and factors 
to realize important features of immune responses.

\end{document}